\documentclass[conference]{IEEEtran}
\IEEEoverridecommandlockouts

\usepackage{cite}
\usepackage{amsmath,amssymb,amsfonts}
\usepackage{algorithm} 
\usepackage{algorithmic}
\usepackage{graphicx}
\usepackage{textcomp}
\usepackage{xcolor}
\usepackage{booktabs}
\usepackage{url}
\usepackage{float} 
\usepackage{tabularx} 
\usepackage{ragged2e} 

\def\BibTeX{{\rm B\kern-.05em{\sc i\kern-.025em b}\kern-.08em
    T\kern-.1667em\lower.7ex\hbox{E}\kern-.125emX}}

\begin{document}

\title{A Critical Review of Monte Carlo Algorithms: \\ Balancing Performance and Probabilistic Accuracy with AI-Augmented Framework }

\author{\IEEEauthorblockN{Ravi Prasad}
\IEEEauthorblockA{\textit{Department of Computer Science \& Engineering} \\
\textit{Mississippi State University}\\
Mississippi State, MS 39762 \\
rp1416@msstate.edu}
}

\maketitle

\begin{abstract}
Monte Carlo algorithms are a foundational pillar of modern computational science, yet their effective application hinges on a deep understanding of their performance trade-offs. This paper presents a critical analysis of the evolution of Monte Carlo algorithms, focusing on the persistent tension between statistical efficiency and computational cost. We describe the historical development from the foundational Metropolis–Hastings algorithm to contemporary methods like Hamiltonian Monte Carlo (HMC). A central emphasis of this survey is the
rigorous discussion of time and space complexity, including upper, lower, and asymptotic tight bounds for each major algorithm class. We examine the specific motivations for developing these methods and the key theoretical and practical observations such as the introduction of gradient information and adaptive tuning in HMC—that led to successively better solutions. Furthermore, we provide a justification framework that discusses explicit situations in which using one algorithm is demonstrably superior to another for the same problem. The paper concludes by assessing the profound significance and impact of these algorithms and detailing major current research challenges.
\end{abstract}


\section{Introduction}

Monte Carlo (MC) algorithms are a broad class of computational algorithms that are indispensable in modern science, engineering, and artificial intelligence. They rely on repeated random sampling to obtain numerical results for problems that are deterministic in principle but too complex to solve analytically. Their applications span numerical integration, optimization, and, most critically for the field of AI, generating samples from high-dimensional and complex probability distributions. This sampling capability forms the backbone of Bayesian inference, where Markov Chain Monte Carlo (MCMC) methods are the primary tool for approximating intractable posterior distributions.

The widespread adoption of MCMC stems from its robust, near-universal applicability. It can, in theory, sample from any target distribution given minimal requirements, a property that makes it a powerful tool. However, this robustness belies a deep and "persistent tension" between two competing objectives: \textit{statistical efficiency} and \textit{computational cost}. Statistical efficiency refers to how well the sampler explores the target distribution, often measured by the Effective Sample Size (ESS) per sample or the autocorrelation between samples. A statistically efficient sampler produces (nearly) independent samples, requiring fewer iterations to achieve a given level of accuracy. Computational cost, conversely, is the measure of time (CPU/GPU cycles) and memory (RAM) required to produce a single sample.

The history of MCMC methods is not a simple linear progression of "improvement," but rather a fascinating study in how this computational burden is transferred. Early algorithms, such as the Random-Walk Metropolis (RWM) algorithm, are computationally \textit{cheap} per iteration (requiring only one likelihood evaluation) but statistically \textit{inefficient}. They exhibit diffusive, random-walk behavior that mixes agonizingly slowly in high-dimensional spaces, thus placing the computational burden on the number of samples $N$ required, which must be enormous. In contrast, contemporary methods like Hamiltonian Monte Carlo (HMC) are statistically \textit{efficient}, leveraging gradient information to propose distant, uncorrelated samples \cite{Hoffman2014, Neal2011}. This reduces the required $N$, but at the cost of a much higher computational burden per sample, as each iteration requires $L$ gradient calculations \cite{Hoffman2014}.

This transfer of the computational burden from the quantity of samples to the complexity of each sample creates a new, critical failure mode: MCMC becomes computationally infeasible if the cost of a single likelihood or gradient evaluation is itself intractable. This is a common scenario in modern scientific AI, where the likelihood may be the output of a large-scale simulator \cite{Zhou2020}.

This paper provides a critical analysis of this evolutionary trade-off and proposes a path forward. The primary contributions are:
\begin{enumerate}
    \item A rigorous analysis of the MCMC evolution from Metropolis to the No-U-Turn Sampler (NUTS), viewed through the lens of this efficiency-cost tension.
    \item A comparative, complexity-theoretic analysis of these algorithms, detailing their time, space, and asymptotic convergence properties.
    \item A formal "justification framework" to guide practitioners in algorithm selection, operationalizing the data from our complexity analysis.
    \item A systematic identification of the persistent "gaps" in the MCMC landscape—namely multi-modality, high-dimensionality, and intractable likelihoods—that remain unsolved by traditional methods.
    \item A novel, modular "AI-Augmented Monte Carlo" (AI-MC) framework, detailing how modern AI—including Deep Generative Models, Reinforcement Learning (RL), and Large Language Models (LLMs)—can be systematically integrated to address these specific gaps, transforming MCMC from a brute-force tool into an intelligent, adaptive agent.
\end{enumerate}

\section{The Historical Evolution of Monte Carlo Algorithm}
\begin{figure*}[t]
    \centering
    \includegraphics[width=\textwidth]{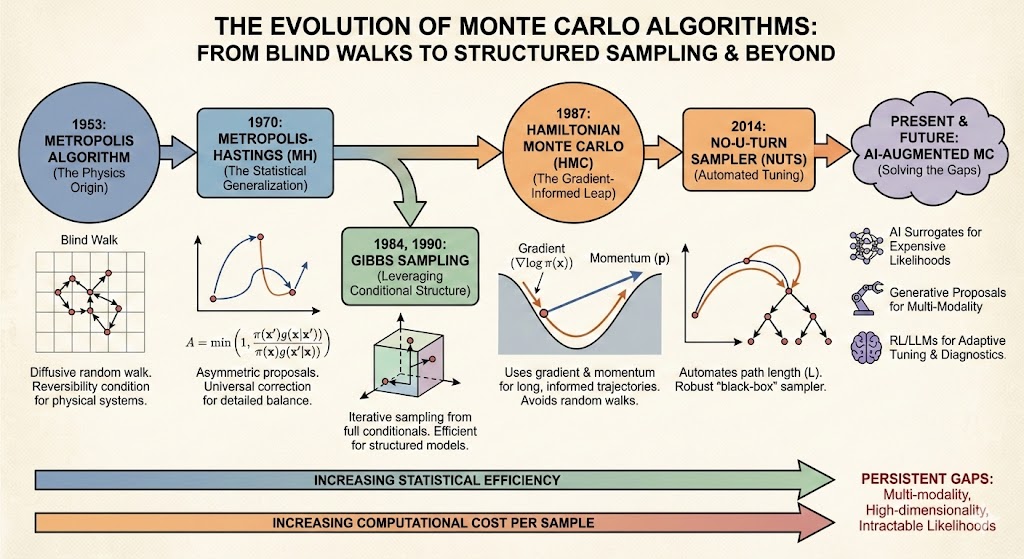}
    \caption{Historical evolution and future direction}
    \label{fig:placeholder}
\end{figure*}

The conceptual- framework for modern MCMC was built in a span of four decades, establishing the theoretical guarantees that underpin all subsequent work. This evolution was driven by key theoretical insights that generalized the algorithms from physics-specific tools to general-purpose statistical engines.

\subsection{The Metropolis Algorithm (1953): A Physics-Based Origin}
The genesis of MCMC is the seminal 1953 paper, "Equation of State Calculations by Fast Computing Machines," by Metropolis, Rosenbluth, Rosenbluth, Teller, and Teller \cite{Metropolis1953}. Developed in the context of statistical physics, its goal was to simulate the distribution of states (e.g., particle positions) for physical systems \cite{Neal2011}.

The core theoretical insight was the mechanism to ensure that the resulting Markov chain would converge to the target (Boltzmann) distribution $\pi(x)$. The algorithm works by proposing a small perturbation to the current state $x$ (e.g., moving a particle) to get a new state $x'$. This proposal is then accepted or rejected based on the change in energy $\Delta E = E(x') - E(x)$, with an acceptance probability $A = \min(1, \exp(-\Delta E/kT))$. This acceptance rule ensures the chain satisfies the \textit{detailed balance} (or \textit{reversibility}) condition, $\pi(x)K(x'|x) = \pi(x')K(x|x')$, which is a sufficient condition for guaranteeing that $\pi$ is the unique stationary distribution. However, the original Metropolis algorithm was limited, as its formulation was designed for \textit{symmetrical proposal distributions} $g(x'|x) = g(x|x')$, where the probability of proposing $x'$ from $x$ is the same as proposing $x$ from $x'$ \cite{Hastings1970}.

\subsection{The Metropolis-Hastings (MH) Algorithm (1970): A Statistical Generalization}
In 1970, W. K. Hastings published "Monte Carlo Sampling Methods Using Markov Chains and Their Applications," \cite{Hastings1970} which generalized the Metropolis algorithm to a vastly broader class of problems. This paper is arguably the true foundation of modern MCMC, as it extended the algorithm to work with \textit{asymmetric proposal distributions} \cite{Hastings1970}.

Hastings' key insight was the formulation of a generalized acceptance probability that includes a "correction factor"—the Hastings ratio—to maintain detailed balance:
$$A(x'|x) = \min\left(1, \frac{\pi(x')g(x|x')}{\pi(x)g(x'|x)}\right)$$
This generalization was a profound theoretical leap. It effectively \textit{decoupled} the problem of \textit{generating} proposals from the problem of \textit{preserving} the stationary distribution. The MH acceptance step acts as a universal correction mechanism; provided the proposal kernel $g(\cdot|\cdot)$ is ergodic (i.e., it can eventually reach any part of the state space), the resulting chain is guaranteed to converge to $\pi(x)$.

This decoupling created a "plug-and-play" framework that defines the entire field of MCMC. All subsequent, complex algorithms—including MALA and HMC—are not alternatives to Metropolis-Hastings. They are, in fact, just highly sophisticated implementations of the MH algorithm, where the entire effort is focused on designing an intelligent proposal distribution $g(x'|x)$ that explores the state space more efficiently \cite{Neal2011}.

\begin{algorithm}[h]
\caption{The Metropolis-Hastings (MH) Algorithm}
\label{alg:mh}
\begin{algorithmic}[1]
\STATE \textbf{Initialize:} starting state $x^{(0)}$, number of samples $N$.
\STATE \textbf{Input:} target distribution $\pi(x)$, proposal kernel $g(x'|x)$.
\FOR{$t = 1$ \algorithmicto $N$}
    \STATE \textit{\# Propose a new state}
    \STATE Sample $x' \sim g(x' | x^{(t-1)})$
    \STATE \textit{\# Calculate acceptance probability}
    \STATE $A(x'|x^{(t-1)}) = \min\left(1, \frac{\pi(x')g(x^{(t-1)}|x')}{\pi(x^{(t-1)})g(x'|x^{(t-1)})}\right)$
    \STATE \textit{\# Accept or reject the proposal}
    \STATE Sample $u \sim U(0, 1)$
    \IF{$u < A(x'|x^{(t-1)})$}
        \STATE $x^{(t)} \leftarrow x'$ \COMMENT{Accept}
    \ELSE
        \STATE $x^{(t)} \leftarrow x^{(t-1)}$ \COMMENT{Reject}
    \ENDIF
\ENDFOR
\STATE \textbf{return} $\{x^{(0)}, x^{(1)},..., x^{(N)}\}$
\end{algorithmic}
\end{algorithm}

\subsection{Gibbs Sampling (1984, 1990): Leveraging Conditional Structure}
While MH provides a general framework, it can be difficult to design a good proposal $g(\cdot|\cdot)$ for a high-dimensional multivariate distribution $p(\theta_1,..., \theta_d | X)$. Gibbs sampling, introduced to statistics by Geman and Geman (1984) and popularized by Gelfand and Smith (1990) \cite{Geman1984, Gelfand1990}, provided a powerful alternative by leveraging the model's structure.

The key insight of Gibbs sampling is to break the high-dimensional problem down into a series of $d$ one-dimensional problems. Instead of sampling the entire vector $\theta$ at once, it iteratively samples each parameter (or block of parameters) from its \textit{full conditional distribution} (FCD), $\theta_i^* \sim p(\theta_i | \theta_{-i}, X)$, using the most recently updated values for all other parameters \cite{Gelfand1990}.

Gibbs sampling can be viewed as a special case of the MH algorithm. The "proposal" for updating $\theta_i$ is simply a draw from its true FCD. When this proposal is plugged into the MH acceptance ratio, the probability of acceptance is exactly 1. This makes Gibbs sampling highly efficient if the FCDs are known and easy to sample from. However, in many complex models, the FCDs are themselves intractable. This leads to the "Metropolis-within-Gibbs" (MwG) algorithm, where a nested MH sampler must be run to update a single parameter. This is computationally costly and highlights the practical limitations that motivated the development of joint-sampling methods like HMC.

\begin{algorithm}[h]
\caption{Gibbs Sampling (for $d$ parameters)}
\label{alg:gibbs}
\begin{algorithmic}[1]
\STATE \textbf{Initialize:} starting state $\mathbf{x}^{(0)} = (x_1^{(0)},..., x_d^{(0)})$, samples $N$.
\FOR{$t = 1$ \algorithmicto $N$}
    \STATE \textit{\# Iterate through each parameter}
    \STATE Sample $x_1^{(t)} \sim \pi(x_1 | x_2^{(t-1)}, x_3^{(t-1)},..., x_d^{(t-1)})$
    \STATE Sample $x_2^{(t)} \sim \pi(x_2 | x_1^{(t)}, x_3^{(t-1)},..., x_d^{(t-1)})$
    \STATE \hspace{0.5cm} $\vdots$
    \STATE Sample $x_j^{(t)} \sim \pi(x_j | x_1^{(t)},..., x_{j-1}^{(t)}, x_{j+1}^{(t-1)},..., x_d^{(t-1)})$
    \STATE \hspace{0.5cm} $\vdots$
    \STATE Sample $x_d^{(t)} \sim \pi(x_d | x_1^{(t)}, x_2^{(t)},..., x_{d-1}^{(t)})$
    \STATE \textit{\# Store the full state for this iteration}
    \STATE $\mathbf{x}^{(t)} \leftarrow (x_1^{(t)}, x_2^{(t)},..., x_d^{(t)})$
\ENDFOR
\STATE \textbf{return} $\{\mathbf{x}^{(0)}, \mathbf{x}^{(1)},..., \mathbf{x}^{(N)}\}$
\end{algorithmic}
\end{algorithm}

\section{Gradient-Informed Sampling: The Hamiltonian Leap}

The foundational "gradient-free" samplers, particularly RWM, suffer from a critical flaw: they explore the state space via simple, diffusive behavior \cite{Neal2011}. This section details the evolution of methods that incorporate geometric information, via gradients, to overcome this limitation.

\subsection{The Inefficiencies of Diffusive Behavior in High Dimensions}
The standard Random-Walk Metropolis (RWM) algorithm, which uses a simple Gaussian proposal $x' \sim N(x, \sigma^2 I)$, is catastrophically inefficient in high-dimensional settings. This is a manifestation of the "curse of dimensionality". To maintain a reasonable acceptance rate (optimally $\sim$23.4\%), the proposal step size $\sigma$ must shrink as $O(1/\sqrt{d})$. This means the number of steps required for the chain to explore the target distribution scales as $O(d)$. The total computational time, therefore, scales poorly.

The problem is not just statistical but geometric. In high dimensions, parameters are often highly correlated, causing the posterior distribution to resemble a long, narrow "valley" or "funnel". A spherical RWM proposal from a point $x$ on this "ridge" will almost certainly propose a new point $x'$ in the "walls" of the valley, leading to a near-100\% rejection rate. The sampler becomes "stuck," exhibiting high autocorrelation and an ESS near zero. This fundamental failure of RWM to handle high-dimensional, correlated posteriors is the \textit{primary motivation} for developing gradient-informed samplers.

\subsection{Metropolis-Adjusted Langevin Algorithm (MALA)}
The first logical step to improve upon RWM is to use local geometry. The Metropolis-Adjusted Langevin Algorithm (MALA) does this by incorporating the gradient of the log-target density, $\nabla \log \pi(x)$. The proposal mechanism is based on a discretization of the Langevin diffusion, which has a "drift term" that pushes proposals "uphill" towards regions of higher probability:
$$x' \sim N(x + \frac{\epsilon^2}{2} \nabla \log \pi(x), \epsilon^2 I)$$
Because this is a discretization, it is not exact, and an MH correction step is required to maintain detailed balance \cite{Robert1996}. MALA is more efficient than RWM, as it uses local information to guide its walk, but it still takes many small, discrete steps and can struggle with complex geometries.

\subsection{Hamiltonian Monte Carlo (HMC): A "Hybrid" Solution}
Hamiltonian Monte Carlo (HMC), originally proposed as "Hybrid Monte Carlo" by Duane et al. (1987) for calculations in lattice quantum chromodynamics \cite{Duane1987, Neal2011}, represents a profound leap in proposal design. It was later adapted for broad use in statistics by Radford Neal \cite{Duane1987, Neal2011}.

The key insight of HMC is to borrow concepts from Hamiltonian dynamics. The $d$-dimensional parameter space $q$ (our "position" variables) is augmented with $d$ fictitious "momentum" variables $p$ \cite{Neal2011}. A Hamiltonian $H(q,p)$ is then defined as the sum of potential energy $U(q)$ and kinetic energy $K(p)$:
$$H(q,p) = U(q) + K(p) = -\log \pi(q) + \frac{1}{2}p^T M^{-1} p$$
where $M$ is a "mass matrix" (typically diagonal) \cite{Hoffman2014}.

The HMC algorithm is an MH algorithm with a remarkably sophisticated proposal mechanism. A single HMC iteration consists of three steps:
\begin{enumerate}
    \item \textbf{Randomize:} The momentum is resampled from its conditional distribution, $p \sim N(0, M)$. This provides the "kick" that starts a new trajectory.
    \item \textbf{Simulate:} The system $(q, p)$ is evolved for a time $T$ by simulating Hamilton's equations of motion. This is done using a \textit{symplectic}, \textit{reversible}, and \textit{volume-preserving} numerical integrator, most commonly the "leapfrog" integrator \cite{Neal2011}. This simulation generates a distant proposal state $(q', p')$.
    \item \textbf{Correct:} This proposal is accepted or rejected using the standard MH rule, based on the change in total energy: $A = \min(1, \exp(-H(q', p') + H(q, p)))$ \cite{Neal2011}.
\end{enumerate}
The genius of HMC lies in this proposal. In an ideal, continuous system, Hamiltonian dynamics perfectly conserves the total energy $H$. This would mean $H(q', p') = H(q, p)$, and the acceptance probability would always be 1 \cite{Neal2011}. Because the leapfrog integrator is a numerical \textit{approximation}, it introduces small errors, so the acceptance step is needed to correct for this discretization \cite{Neal2011}. However, the approximation is extremely accurate, so the acceptance rate remains very high (near 1) even for proposals $q'$ that are \textit{very far} from $q$.

\begin{algorithm}[h]
\caption{The Leapfrog Integrator (Sub-procedure)}
\label{alg:leapfrog}
\begin{algorithmic}[1]
\STATE{Leapfrog}{$(q, p, \epsilon, L, \nabla U$)}
\STATE $p \leftarrow p - \frac{\epsilon}{2} \nabla U(q)$ \COMMENT{Half-step for momentum}
\FOR{$i = 1$ \algorithmicto $L$}
    \STATE $q \leftarrow q + \epsilon M^{-1} p$ \COMMENT{Full-step for position}
    \IF{$i \neq L$}
        \STATE $p \leftarrow p - \epsilon \nabla U(q)$ \COMMENT{Full-step for momentum}
    \ENDIF
\ENDFOR
\STATE $p \leftarrow p - \frac{\epsilon}{2} \nabla U(q)$ \COMMENT{Final half-step for momentum}
\STATE \algorithmicreturn $(q, p)$
\end{algorithmic}
\end{algorithm}

\begin{algorithm}[h]
\caption{Hamiltonian Monte Carlo (HMC) Algorithm}
\label{alg:hmc}
\begin{algorithmic}[1]
\STATE \textbf{Initialize:} starting state $q^{(0)}$, samples $N$.
\STATE \textbf{Input:} step size $\epsilon$, steps $L$, mass matrix $M$,
\STATE \hspace{1.4cm} potential energy $U(q) = -\log \pi(q)$.
\FOR{$t = 1$ \algorithmicto $N$}
    \STATE \textit{\# 1. Randomize momentum}
    \STATE Sample $p^{(0)} \sim N(0, M)$
    \STATE Set $(q, p) \leftarrow (q^{(t-1)}, p^{(0)})$
    \STATE \textit{\# 2. Simulate trajectory}
    \STATE $(q', p') \leftarrow \text{Leapfrog}((q, p), \epsilon, L, \nabla U)$
    \STATE \textit{\# 3. Metropolis-Hastings Correction}
    \STATE $H(q, p) = U(q) + \frac{1}{2} p^T M^{-1} p$
    \STATE $\alpha = \min\left(1, \exp\left(H(q, p) - H(q', -p')\right)\right)$
    \STATE \textit{\# Note: momentum is flipped for reversibility test}
    \STATE Sample $u \sim U(0, 1)$
    \IF{$u < \alpha$}
        \STATE $q^{(t)} \leftarrow q'$ \COMMENT{Accept}
    \ELSE
        \STATE $q^{(t)} \leftarrow q^{(t-1)}$ \COMMENT{Reject}
    \ENDIF
\ENDFOR
\STATE \textbf{return} $\{q^{(0)}, q^{(1)},..., q^{(N)}\}$
\end{algorithmic}
\end{algorithm}

HMC thus "solves" the central trade-off of RWM. It makes large, accepted moves, allowing it to rapidly explore the parameter space and avoid the slow, diffusive random walk \cite{Neal2011, Hoffman2014}.

\subsection{From HMC to NUTS: Automating the "Tuning Nightmare"}
The power of HMC comes at a cost: it introduces new, highly sensitive hyperparameters, namely the leapfrog step size $\epsilon$ and the number of steps $L$ (which defines the total trajectory length $T = \epsilon L$) \cite{Hoffman2014}. Manually tuning these parameters is notoriously difficult. If $L$ is too short, HMC degenerates into a MALA-like random walk. If $L$ is too long, the trajectory follows the curvature of the posterior, "makes a U-turn," and starts to come back on itself, wasting all subsequent computational effort \cite{Hoffman2014}.

The No-U-Turn Sampler (NUTS), introduced by Hoffman and Gelman in 2014, was designed to solve this problem by automating the selection of $L$* \cite{Hoffman2014}. NUTS runs the leapfrog integrator in both forward and backward time, building a binary tree of states. It automatically stops the simulation when a "U-turn" is detected—that is, when continuing the trajectory would decrease the distance between the proposal and the initial state \cite{Hoffman2014}. It then samples a new state from the set of valid points in the generated tree, using a mechanism that preserves detailed balance.

NUTS was the key innovation that transformed HMC from an expert-only tool into a robust, "black-box" sampler \cite{Hoffman2014}. This automation is why HMC (in the form of NUTS) is the default sampling engine in modern Bayesian software like Stan \cite{Neal2011, Hoffman2014}. It is not without its own trade-offs; NUTS's tree-building and backtracking mechanism "wastes about half of its leapfrog steps" to satisfy detailed balance \cite{Hoffman2014}. This means that a perfectly, manually-tuned HMC with a fixed $L$ can be more computationally efficient, but NUTS provides the robustness and automation required for general-purpose use.

\section{A Comparative Analysis of Complexity and Efficiency}

To formalize the trade-offs discussed, this section provides a rigorous comparison of the major MCMC algorithms. This analysis forms the quantitative basis for the justification framework in Section V.

\subsection{Defining the Axes of Comparison}
We evaluate the algorithms along two primary axes: computational cost and statistical efficiency.

\subsubsection{Computational Cost (Time and Space)}
\begin{itemize}
    \item \textbf{Time Complexity:} Measured as the wall-clock time required per MCMC iteration. We define $d$ as the dimension of the parameter space, $C_f$ as the computational cost of evaluating the log-target density $\log \pi(x)$, and $C_g$ as the cost of evaluating its gradient $\nabla \log \pi(x)$. Using modern automatic differentiation (AD), $C_g$ is typically a small constant multiple of $C_f$, (e.g., $C_g \approx 3 \cdot C_f$), and scales with the complexity of the model, which is often at least $O(d)$ \cite{Neal2011}.
    \item \textbf{Space Complexity:} Measured as the memory (RAM) required to store the chain's state and any auxiliary information needed for a single iteration.
\end{itemize}

\subsubsection{Statistical Efficiency (Convergence and Mixing)}
\begin{itemize}
    \item \textbf{Asymptotic Bounds:} Theoretically, we are concerned with the mixing time of the Markov chain (i.e., how fast it converges to $\pi(x)$) and the asymptotic variance of the MCMC estimator, $\hat{\theta}_n = \frac{1}{n} \sum f(X_i)$. This variance is given by $\sigma^2_{as}(P,f)/n$. A more statistically efficient sampler is one that produces a smaller $\sigma^2_{as}$, as this means fewer samples $n$ are required for a given error tolerance.
    \item \textbf{Practical Metrics:} In practice, statistical efficiency is measured by the \textbf{Effective Sample Size (ESS)} and the \textbf{autocorrelation} of the samples. A sampler with low autocorrelation produces a high ESS per sample, rapidly exploring the target distribution. HMC and NUTS are explicitly designed to reduce autocorrelation compared to RWM \cite{Neal2011}.
\end{itemize}

\subsection{Algorithm Complexity Comparison}
The properties of the major algorithm classes are summarized in Table I.

\begin{itemize}
    \item \textbf{Random-Walk Metropolis (RWM):} This is the baseline. Its time cost per iteration is minimal, $O(C_f)$, and its space cost is $O(d)$. However, its statistical efficiency is extremely poor. The diffusive behavior means its mixing time scales as $O(d^2)$ or worse \cite{Neal2011}.
    \item \textbf{Gibbs Sampling:} Complexity is highly problem-dependent. Time cost is $O(\sum_{i=1}^d C_{FCD_i})$, where $C_{FCD_i}$ is the cost of sampling from the $i$-th full conditional. If parameters are weakly correlated and FCDs are simple, it can be the most efficient method. If parameters are highly correlated, its performance degrades and can be as bad as RWM's \cite{Gelfand1990}.
    \item \textbf{Hamiltonian Monte Carlo (HMC):} Here the trade-off is explicit. For a fixed number of leapfrog steps $L$, the time cost per iteration is $O(L \cdot C_g)$ \cite{Hoffman2014}. This is a \textit{significant} upfront computational cost \cite{Hoffman2014}. The benefit is a massive gain in statistical efficiency. HMC avoids the $O(d^2)$ random walk; its mixing time can scale as well as $O(d)$ or, in some ideal cases, $O(d^{1/4})$ \cite{Hoffman2014, Neal2011}.
    \item \textbf{No-U-Turn Sampler (NUTS):} NUTS inherits the statistical efficiency of HMC. Its time cost is $O(L' \cdot C_g)$, where $L'$ is the variable number of steps determined by the tree-building process \cite{Hoffman2014}. This introduces a space complexity trade-off: NUTS must store the entire binary tree of states in memory during a single iteration, leading to a space cost of $O(L' \cdot d)$, which can be non-trivial for long trajectories. HMC, by contrast, only needs $O(d)$ space \cite{Hoffman2014}.
\end{itemize}

\begin{table*}[t]
\caption{Comparative Analysis of MCMC Algorithm Properties}
\label{tab:complexity}
\centering
\begin{tabularx}{\textwidth}{@{} l >{\RaggedRight}X l >{\RaggedRight}X >{\RaggedRight}X >{\RaggedRight}X @{}}
\toprule
\textbf{Algorithm} & \textbf{Core Proposal Mechanism} & \textbf{Requires Gradient?} & \textbf{Time Complexity (per iter, $d$ dim, $L$ steps)} & \textbf{Space Complexity (per iter)} & \textbf{Statistical Efficiency (Mixing Time)} \\
\midrule
\textbf{RWM} & $q' \sim N(q, \sigma^2 I)$ & No & $O(C_f)$ & $O(d)$ & Poor \cite{Neal2011}. $O(d^2)$ or worse. \\
\textbf{Gibbs} & $q_i \sim \pi(q_i | q_{-i})$ & No & $O(\sum C_{FCD_i})$ & $O(d)$ & Varies. Can be $O(1)$ or $O(d^2)$ depending on correlation \cite{Gelfand1990}. \\
\textbf{MALA} & Discretized Langevin step & Yes & $O(C_g)$ & $O(d)$ & Good. Better than RWM \cite{Robert1996}. \\
\textbf{HMC} & Symplectic integrator (fixed $L$) & Yes & $O(L \cdot C_g)$ & $O(d)$ & Excellent \cite{Hoffman2014}. $O(d)$ to $O(d^{1/4})$. \\
\textbf{NUTS} & Symplectic integrator (dynamic $L'$) & Yes & $O(L' \cdot C_g)$ & $O(L' \cdot d)$ & Excellent \cite{Hoffman2014}. $O(d)$ to $O(d^{1/4})$. \\
\bottomrule
\end{tabularx}
\end{table*}

\section{A Justification Framework for Algorithm Selection}

The complexity analysis in Section IV provides the data to construct a practical, decision-theoretic framework for practitioners. This framework, presented as a decision tree, formalizes the answer to the question: "Which MCMC algorithm should I use for my problem?"
\begin{figure}
    \centering
    \includegraphics[width=1\linewidth]{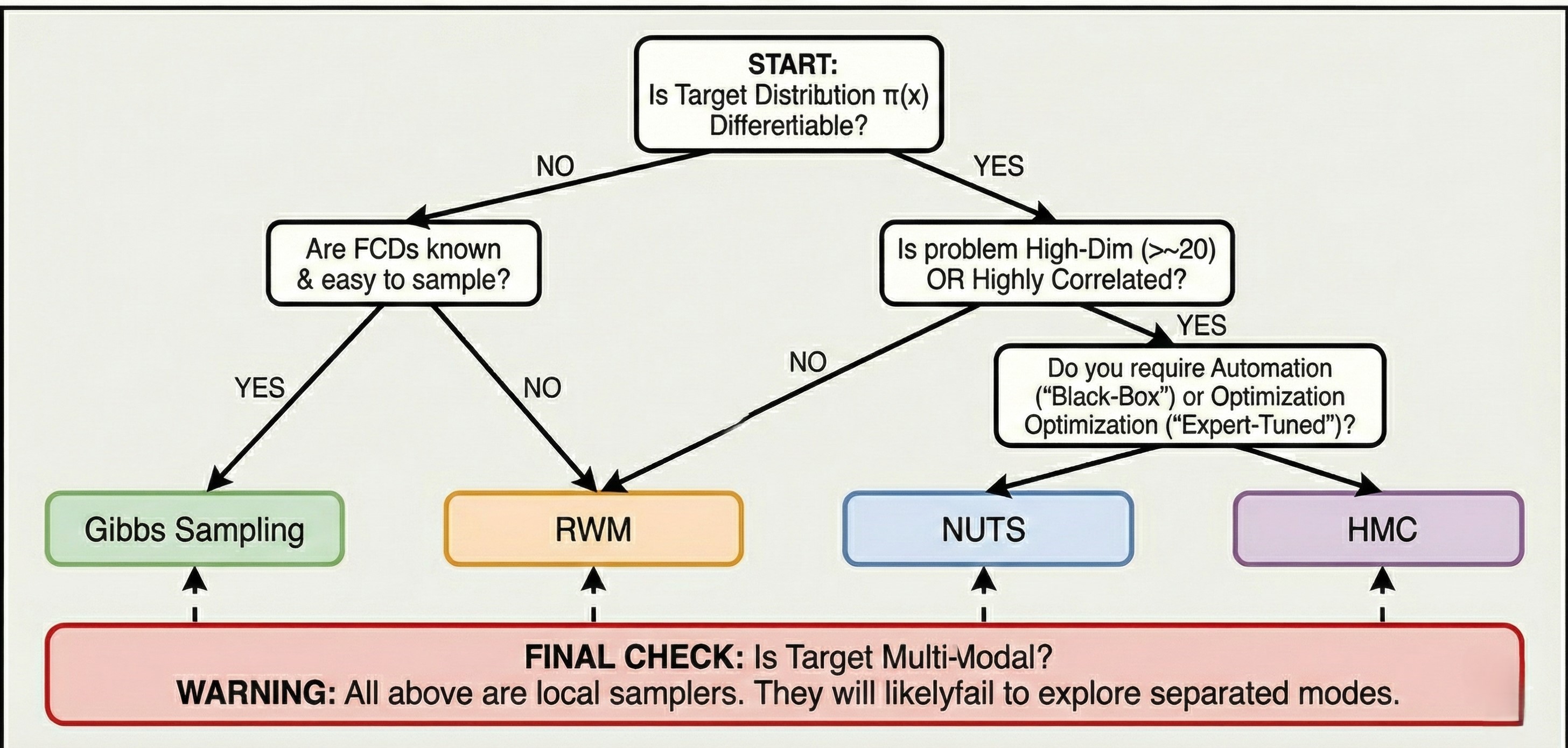}
    \caption{A Framework for algorithm selection}
    \label{fig:placeholder}
\end{figure}

\subsection{Decision Point 1: Differentiability of the Target}
The first and most important question is: \textbf{Is the target distribution $\pi(x)$ differentiable?}
\begin{itemize}
    \item \textbf{NO:} If the log-target $\log \pi(x)$ is not differentiable (e.g., the model involves discrete parameters, black-box functions, or discontinuous steps), then gradient-based methods are not applicable. You cannot use MALA, HMC, or NUTS.
    \item \textbf{PROCEED TO: Gradient-Free Path.}
    \item \textbf{YES:} If $\nabla \log \pi(x)$ can be computed, either analytically or via automatic differentiation (e.g., in standard Bayesian models, neural networks), then the powerful gradient-based samplers are available.
    \item \textbf{PROCEED TO: Gradient-Based Path.}
\end{itemize}

\subsection{Gradient-Free Path: Conditional Structure}
Given a non-differentiable target, the next question concerns its conditional structure: \textbf{Are the Full Conditional Distributions (FCDs) known and easy to sample from?}
\begin{itemize}
    \item \textbf{YES:} \textbf{Use Gibbs Sampling.} If the FCDs are tractable (e.g., in conjugate models), Gibbs sampling is almost always the best choice \cite{Gelfand1990}. It is exact (acceptance rate is 1), computationally simple, and can be very efficient.
    \item \textbf{NO:} \textbf{Use Random-Walk Metropolis (RWM)} or another gradient-free sampler (e.g., Slice Sampling). This is the "fallback" algorithm. Be aware that it will suffer from the curse of dimensionality and mix very slowly if $d$ is high or parameters are correlated.
\end{itemize}

\subsection{Gradient-Based Path: Dimensionality and Correlation}
Given a differentiable target, the choice to use gradients depends on the problem's geometry: \textbf{Is the problem high-dimensional ($d > \sim 20$) AND/OR are parameters highly correlated?}
\begin{itemize}
    \item \textbf{NO:} For a simple, low-dimensional problem with weak correlations, \textbf{RWM is likely sufficient.} The computational overhead of calculating gradients ($C_g$) may not be worth the statistical gain.
    \item \textbf{YES:} This is the \textit{exact} scenario where RWM fails catastrophically. The gradients are no longer optional; they are \textit{required} to efficiently navigate the complex, high-dimensional posterior.
\end{itemize}

\subsection{Gradient-Based Path: Automation vs. Optimization}
Given that a gradient-based sampler is required, the final choice is between automation and expert-tuning: \textbf{Do you require a "black-box" sampler or are you optimizing a fixed model for maximum performance?}
\begin{itemize}
    \item \textbf{"Black-Box" (Automation):} \textbf{Use NUTS.} It is the state-of-the-art, automated gradient sampler that adaptively tunes its trajectory length \cite{Hoffman2014}. It is the robust, default choice in modern software like Stan and is highly recommended for general-purpose use \cite{Neal2011, Hoffman2014}.
    \item \textbf{"Expert-Tuned" (Optimization):} \textbf{Use HMC.} If you are an expert and your goal is to squeeze maximum performance from a single, fixed model (e.g., in a production environment), a manually and optimally-tuned HMC with fixed $\epsilon$ and $L$ may achieve a higher ESS-per-second than NUTS, as it avoids the computational overhead of tree-building \cite{Hoffman2014}.
\end{itemize}

\subsection{Final Check: The Unaddressed Gap (Multi-Modality)}
This framework, while comprehensive for standard problems, has a critical blind spot. A final check is required: \textbf{Is the target distribution $\pi(x)$ known or suspected to be multi-modal?}
\begin{itemize}
    \item \textbf{YES: WARNING.} \textit{All} algorithms in this framework (RWM, Gibbs, HMC, NUTS) are \textit{local} samplers. They are designed to efficiently explore a single, connected high-probability region. If the distribution has multiple, well-separated modes ("deep valleys"), these samplers will almost certainly \textit{fail}. They will get trapped in the first mode they find and will be practically unable to "jump" to the others. This failure to sample from multi-modal distributions is not a minor issue; it is a persistent, fundamental gap in the MCMC landscape, which this framework reveals.
\end{itemize}

\section{Persistent Gaps in the MCMC Landscape}

The justification framework in Section V demonstrates that while modern MCMC methods are powerful, they are not panaceas. They are vulnerable to specific, well-known "gaps" or "failure modes" where even the most advanced samplers, like NUTS, become computationally infeasible or produce incorrect, biased results. Our proposed AI-MC framework is designed to specifically address these gaps.

\subsection{Gap 1: The Challenge of Multi-Modality}
This is the most critical gap for correctness. As noted, standard MCMC algorithms are local explorers. They follow a path of non-zero probability. To move from one high-probability mode to another, they must traverse the low-probability "valley" or "energy barrier" separating them. The time required to make such a jump is often exponential in the height of this barrier. In practice, this means the chain will spend its entire runtime trapped in a single mode, leading to posterior estimates that are completely non-ergodic and severely biased, missing entire regions of the state space. HMC, with its gradient-following behavior, can be \textit{even more} susceptible to this trapping than a simple RWM sampler.

\subsection{Gap 2: The Curse of Dimensionality}
While HMC/NUTS are vastly superior to RWM in high dimensions (e.g., $O(d)$ vs $O(d^2)$ mixing), they do not "solve" the curse of dimensionality. Their cost per iteration, $O(L' \cdot C_g)$, still scales with the dimension. The cost of the gradient calculation $C_g$ is often at least $O(d)$, and for models with dense interactions (like a large covariance matrix), $C_g$ could be $O(d^2)$. For problems in modern AI, such as Bayesian neural networks, $d$ can be in the millions or billions. In this regime, even the $O(d)$ scaling of HMC is computationally prohibitive, and the $O(L' \cdot d)$ \textit{space} complexity of NUTS can exceed available memory.

\subsection{Gap 3: The Problem of Expensive or Intractable Likelihoods}
All algorithms discussed, from RWM to NUTS, share a fatal assumption: that evaluating the log-likelihood $\log \pi(x)$ (and its gradient) is computationally cheap. In many cutting-edge scientific domains (e.g., climate modeling, computational biology, particle physics, subsurface remediation), this assumption is false \cite{Zhou2020, Jarvenpaa2024}. The likelihood function is not a simple equation; it is the \textit{output of a complex, computationally-intensive simulator} (e.g., a finite element model or a system of differential equations) \cite{Zhou2020}. A single evaluation of $\log \pi(x)$ might take minutes, hours, or even days. In this "expensive likelihood" scenario, an MCMC run of 10,000 iterations is a computational impossibility. An HMC/NUTS iteration, which requires $L$ gradient calls (each as expensive as $C_f$), is even more infeasible.

\subsection{Gap 4: The Hyperparameter Tuning "Nightmare"}
This is a "meta-problem." The performance of advanced samplers is critically sensitive to their hyperparameters. For HMC/NUTS, these include the leapfrog step size $\epsilon$ and the mass matrix $M$ \cite{Hoffman2014}. Finding optimal values for these is a difficult, high-dimensional optimization problem in its own right \cite{Snoek2012}. A poor choice of $\epsilon$ can lead to high rejection rates or numerical instability. A poor choice of $M$ (which should approximate the posterior covariance) can lead to the same "narrow valley" problem that HMC was meant to solve. This tuning process is often described as a "dark art" that barriers practical application and wastes significant researcher and computation time.

\section{A Framework for AI-Augmented Monte Carlo (AI-MC)}

We propose a novel, modular framework to systematically address the gaps identified in Section VI. The philosophy of this AI-Augmented Monte Carlo (AI-MC) framework is to create a \textit{synergistic hybrid system}. This system retains the \textit{statistical core} of MCMC (specifically, the Metropolis-Hastings acceptance step) to ensure theoretical guarantees of convergence, while integrating \textit{intelligent, learnable AI components} to handle the computationally difficult tasks of proposal generation, likelihood evaluation, and hyperparameter tuning.
\begin{figure}[h]
    \centering
    \includegraphics[width=1\linewidth]{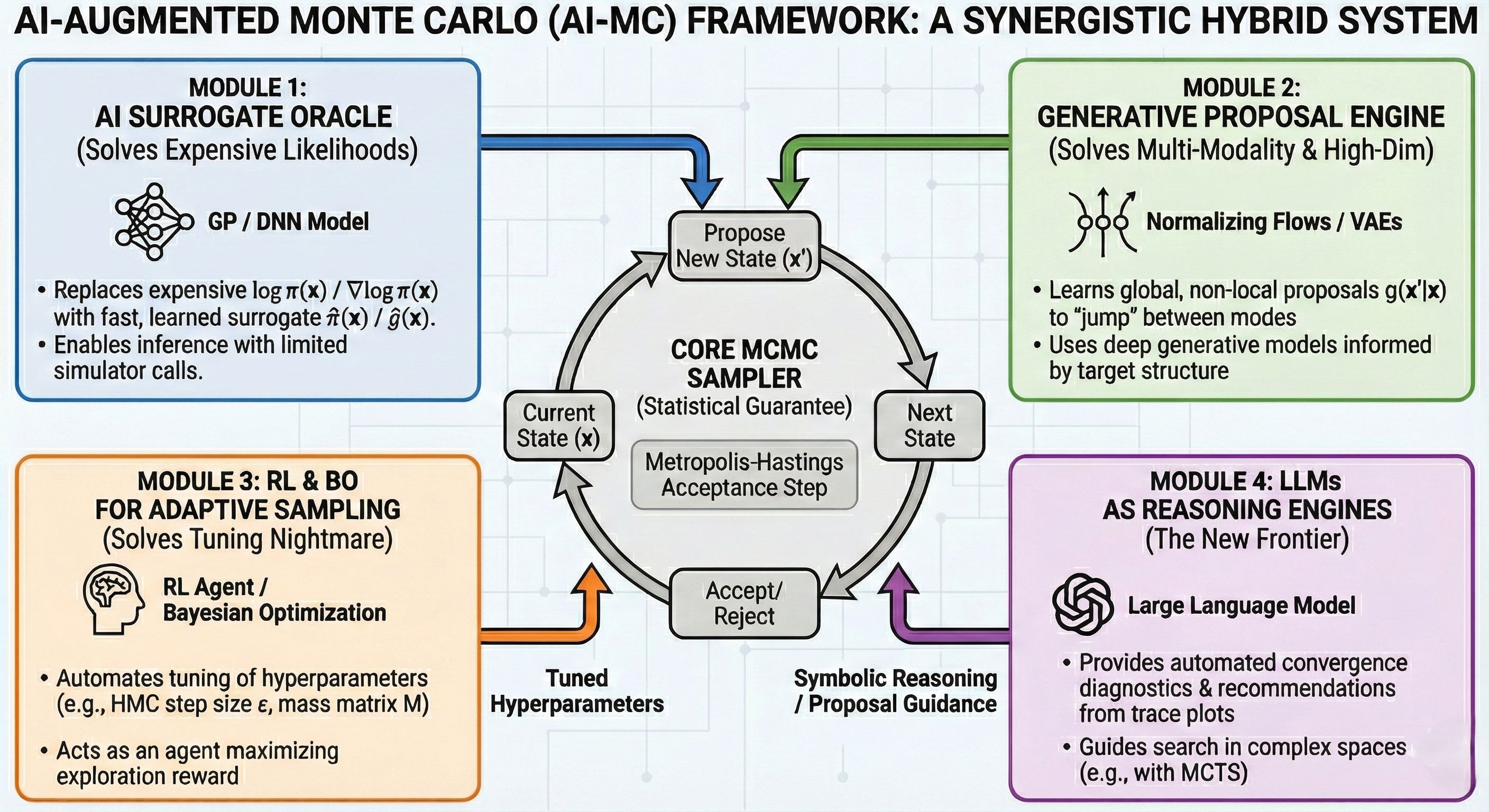}
    \caption{A Framework for AI-Augmented Monte Carlo (AI-MC)}
    \label{fig:placeholder}
\end{figure}
\subsection{Module 1: AI as a Surrogate Oracle (Solves Gap 3: Expensive Likelihoods)}
To solve the intractable likelihood problem, we replace the expensive "true" function $\log \pi(x)$ or its gradient $\nabla \log \pi(x)$ with a fast, learned \textit{surrogate model} $\hat{\pi}(x)$ or $\hat{g}(x)$.

\begin{itemize}
    \item \textbf{Method 1: Gaussian Process (GP) Surrogates.} For problems where every likelihood evaluation is precious, a GP can be used to model $\log \pi(x)$ \cite{Jarvenpaa2024}. This approach is ideal for \textit{active learning} or \textit{Bayesian experimental design}, where the GP's uncertainty estimate is used to intelligently select the \textit{next} parameter $\theta$ at which to run the expensive simulator, maximizing the information gain \cite{Jarvenpaa2024}. This "GP-emulated MCMC" can perform approximate inference with a very limited number of simulator calls \cite{Jarvenpaa2024}.
    \item \textbf{Method 2: Deep Neural Network (DNN) Surrogates.} For problems where an initial "training budget" of simulator runs is feasible, a DNN (e.g., a CNN or a neural operator) can be trained to learn the mapping from parameters $\theta$ to the simulator's output \cite{Zhou2020}. This highly accurate surrogate is then "plugged into" the MCMC sampler \cite{Zhou2020}. The sampler (e.g., HMC) can then request millions of \textit{virtual} likelihood or gradient evaluations from the DNN in seconds, enabling posterior inference for complex, simulator-based models that were previously intractable \cite{Zhou2020}.
\end{itemize}

\subsection{Module 2: Generative Models as Proposal Engines (Solves Gaps 1 \& 2)}
To solve multi-modality (Gap 1), we need a \textit{global, non-local} proposal $g(x'|x)$ that can "jump" between modes. To solve high-dimensional geometry (Gap 2), this proposal must be \textit{informed} by the target's complex structure. This is precisely the function of deep generative models \cite{Albergo2021}.

\begin{itemize}
    \item \textbf{Method 1: Normalizing Flows (NFs).} NFs are a class of generative models that learn an explicit, \textit{bijective mapping} $f$ from a simple base distribution (e.g., $z \sim N(0,I)$) to a complex target distribution $\pi(x)$ \cite{Gabrie2022}. An NF can be trained \textit{adaptively} during the MCMC run to approximate $\pi(x)$ \cite{Gabrie2022}. Once trained, this flow $f$ acts as a perfect "Neural Transport" mechanism \cite{Gabrie2022}:
        \begin{enumerate}
            \item Draw a simple latent sample $z' \sim N(0,I)$.
            \item Propose $x' = f(z')$.
        \end{enumerate}
    This $x'$ is an (approximate) independent draw from $\pi(x)$, allowing the chain to "teleport" from one mode to another in a single MH step, thus solving the multi-modal trapping problem \cite{Gabrie2022}.
    \item \textbf{Method 2: Variational Autoencoders (VAEs).} VAEs can also be used to learn a compressed, latent representation of a multi-modal distribution. The VAE's generative decoder can be used as a proposal mechanism, or the MCMC can be run in the simpler, disentangled latent space, which is then mapped back to the data space.
\end{itemize}

\subsection{Module 3: Reinforcement Learning \& BO for Adaptive Sampling (Solves Gap 4)}
To solve the hyperparameter "tuning nightmare," we can automate the process by framing it as a formal optimization or control problem.

\begin{itemize}
    \item \textbf{Method 1: Bayesian Optimization (BO).} The MCMC hyperparameters (e.g., HMC step size $\epsilon$, mass matrix $M$) can be treated as inputs to a "black-box" objective function, $f(\theta) = \text{ESS(MCMC-run with } \theta \text{)}$, where ESS is the effective sample size \cite{Snoek2012}. BO can be used to \textit{intelligently} search for the $\theta$ that maximizes this objective. It does so by building a surrogate model (typically a GP) of the performance landscape, allowing it to efficiently balance exploration and exploitation \cite{Snoek2012}. This approach has been shown to find hyperparameter configurations that outperform human experts \cite{Snoek2012}.
    \item \textbf{Method 2: Reinforcement Learning (RL).} The MCMC sampler can be framed as an "agent" whose goal is to explore the state space \cite{Wang2024}. The "state" is the current position $x$ of the chain, the "action" is the choice of MCMC kernel parameters (e.g., proposal variance, HMC trajectory length), and the "reward" is a metric of exploration (e.g., squared jump distance). An RL agent can learn a \textit{policy} $\pi(a|s)$ that dynamically adapts the sampler's parameters "on the fly" to maximize the cumulative reward, leading to a chain that automatically adjusts its behavior to the local geometry of the distribution \cite{Wang2024}.
\end{itemize}

\subsection{Module 4: LLMs as Reasoning and Diagnostic Engines (The New Frontier)}
This is the most forward-looking module, leveraging the unique capabilities of Large Language Models (LLMs) not just as pattern recognizers, but as \textit{reasoning engines}.

\begin{itemize}
    \item \textbf{Application 1: Automated Convergence Diagnostics.} In practice, MCMC convergence is assessed by a human "in the loop" visually inspecting trace plots or diagnostic statistics. An LLM can be trained to "read" the output of standard diagnostics (e.g., trace plots, Gelman-Rubin $\hat{R}$ statistics \cite{Gelfand1990}) and provide a natural language summary and recommendation, e.g., "Warning: The $\hat{R}$ for parameter '$\beta\_1$' is 1.4. The trace plot shows the chain is non-stationary and has not converged. Recommend increasing warmup iterations or re-parameterizing the model.".
    \item \textbf{Application 2: LLM-Guided Search and Proposal.} For sampling on complex, structured spaces (e.g., program synthesis, molecular design), an LLM can be combined with Monte Carlo Tree Search (MCTS). The LLM acts as the "policy" (guiding which action to take) and "value" (estimating the reward) functions, allowing the MCTS to intelligently prune unpromising search branches.
    \item \textbf{Application 3: LLM as a Generative Prior/Proposal.} In a highly exploratory approach, an LLM can be used \textit{directly} as a generative proposal mechanism $g(x'|x)$. By prompting the LLM with the chain's history (e.g., "The last 10 accepted states were..."), it can be asked to "reason" about a good next step. This can be seen as a form of in-context Bayesian inference, effectively integrating symbolic reasoning into the statistical sampling loop.
\end{itemize}

\section{Future Directions and Major Research Challenges}

The AI-MC framework, while promising, is not a final solution. It opens up an entirely new set of research challenges centered on the interface between black-box AI and statistically rigorous sampling.

\subsection{Challenge 1: Quantifying and Controlling Surrogate Error}
The central problem of the AI-MC framework is \textit{trust}. If the AI surrogate (Module 1) or generative proposal (Module 2) is inaccurate, it will introduce a bias into the MCMC chain. While the MH correction step prevents the chain from converging to the wrong distribution, a poor surrogate will lead to a near-zero acceptance rate, causing the chain to stop mixing entirely. The key research challenge is to develop a "unified framework" for approximate MCMC methods that provides non-asymptotic bounds on the distance between the
approximate chain and the true posterior. How much error does a 99\% accurate surrogate introduce? How do we detect and control this?

\subsection{Challenge 2: Theoretical Guarantees for Adaptive AI-MC}
Standard MCMC convergence proofs rely on the Markov property. An AI-MC sampler, however, is non-Markovian. A sampler guided by an adaptive RL agent (Module 3) or a stateful LLM (Module 4) has a transition kernel that changes at every step based on the entire history of the chain. This requires extending the theory of adaptive MCMC \cite{Gabrie2022} to these highly complex, non-linear, and high-dimensional "helpers." We must establish formal conditions, such as "diminishing adaptation" \cite{Wang2024}, to prove that these intelligent agents eventually stop "learning" and allow the chain to converge to its true stationary distribution.

\subsection{Challenge 3: The "Data Quality" Vicious Cycle}
The AI modules (especially surrogates and generative proposals) are trained on data. But in MCMC, this data is generated by the MCMC chain itself. This creates a dangerous potential for a "vicious cycle." If the MCMC chain starts in a bad state (e.g., trapped in a local mode), it will generate poor-quality training data. This data, when used to train the AI module, will result in a poor surrogate or proposal engine. This bad AI module will then reinforce the chain's bad behavior, trapping it permanently. Breaking this feedback loop is a critical research challenge.

\subsection{Challenge 4: Computational Overhead and Amortization}
The AI-MC framework is not "free." Training a large DNN surrogate \cite{Zhou2020}, a Normalizing Flow \cite{Gabrie2022}, or an RL agent \cite{Wang2024} is computationally expensive. The new research challenge is to move from an "cost-per-sample" analysis to a holistic "total-budget" analysis. Is it worth spending 24 hours training an AI surrogate if it reduces a 200-hour MCMC run to just 1 hour? This amortized cost-benefit analysis will be essential for justifying these methods and requires a new way of thinking about the computational budget.

\section{Conclusion}

This paper has traced the historical evolution of Monte Carlo methods, from the foundational Metropolis-Hastings algorithm \cite{Metropolis1953, Hastings1970} to the sophisticated, gradient-driven Hamiltonian Monte Carlo \cite{Neal2011, Duane1987} and its automated variant, NUTS \cite{Hoffman2014}. We have shown that this evolution is defined by a "persistent tension" between statistical efficiency and computational cost, a trade-off that has been navigated by transferring the computational burden from the number of samples to the complexity of each sample.

Through a rigorous complexity analysis (Table I) and a formal "justification framework," we demonstrated that this evolution, while successful, has left several persistent "gaps" in the MCMC toolbox. Even the most advanced traditional samplers fail when faced with multi-modal distributions, extreme dimensionality, computationally intractable likelihoods \cite{Jarvenpaa2024}, and complex hyperparameter tuning spaces.

The novel AI-Augmented Monte Carlo (AI-MC) framework proposed here offers a systematic path forward. By integrating AI as intelligent, learnable components—such as DNN surrogates for intractable likelihoods \cite{Zhou2020}, generative models for global proposals \cite{Albergo2021, Gabrie2022}, and RL agents for automated tuning \cite{Snoek2012, Wang2024}—we can plug these gaps directly. This framework even extends to leveraging the reasoning capabilities of LLMs for diagnostics and proposal generation.

The future of MCMC is not a replacement by AI, but a \textit{synergistic hybrid}. MCMC provides the rigorous statistical framework for guarantees and error control, while AI provides the adaptive, geometry-aware, and reasoning-based intelligence required to scale these methods. This synthesis will be essential for tackling the next generation of complex, high-dimensional, and computationally massive inference problems across science and artificial intelligence.

\bibliographystyle{IEEEtran}
\bibliography{biblography}
\nocite{*}
\end{document}